\documentclass[aps,showpacs]{revtex4}
\usepackage{graphicx}
\usepackage{amsmath}
\usepackage{amssymb}
\usepackage{amsfonts}

\newcommand{\avg}[1]{\left<{#1}\right>}

\newcommand{\fkn}{\dfrac{\omega_k}{2\pi}}
\newcommand{\lbr}{\left(}
\newcommand{\rbr}{\right)}

\begin{document}
\title{$1/f$ noise in experimental Sinai quantum billiards}

\author{
E. Faleiro$^{1}$, U. Kuhl$^{2}$, R. A. Molina$^{3}$, A. Rela\~no$^{4}$,
J. Retamosa$^{4}$, and H.-J. St\"ockmann$^{2}$}

\address{$^{1}$ Departamento de F\'{\i}sica Aplicada,E. U. I. T. Industrial, \\
Universidad Polit\'ecnica de Madrid, E-28012 Madrid, Spain \\
$^{2}$ Fachbereich Physik, Philipps-Universit\"at, Renthof 5,\\
D-35032 Marburg, Germany \\
$^{3}$ Max-Planck-Institut f\"ur Physik Komplexer Systeme \\
N\"othnitzer Str. 38, D-01187 Dresden, Germany\\
$^{4}$ Departamento de F\'{\i}sica At\'omica, Molecular y Nuclear,\\
Universidad Complutense de Madrid, E-28040 Madrid, Spain
          }




\begin{abstract}

It was recently conjectured that {\em 1/f} noise is a fundamental
characteristic of spectral fluctuations in chaotic quantum systems. In
this Letter we show that the level fluctuations of experimental
realizations of the Sinai billiard exhibit {\em 1/f} noise,
corroborating the conjecture.  Assuming that the statistical
properties of these systems are those of the Gaussian orthogonal
ensemble (GOE), we compare the experimental results with the universal
behavior predicted in the random matrix framework, and find an
excellent agreement. The deviations from this behavior observed at low
frequencies can be easily explained with the semiclassical periodic
orbit theory. In conclusion, it is shown that the main features of the
conjecture are affected neither by non-universal properties due to the
underlying classical dynamics nor by the uncertainties of the
experimental setup.

\end{abstract}


\maketitle

%
%

Quantum chaos is a well established subject since the pioneering work
of Bohigas and collaborators \cite{Bohigas:84} (for further review
see, for example, \cite{Guhr:98,Stockman:99}). Recently, a different
approach to quantum chaos has been proposed
\cite{Relano:02}. Considering the sequence of energy levels as a
discrete time series, and the appropriate statistic related to the
fluctuations of the excitation energy, it was conjectured that chaotic
quantum systems are characterized by $1/f$ noise, whereas integrable
ones exhibit $1/f^2$ noise. Since chaotic spectra behave like
antipersistent time series, and regular ones as neither persistent nor
antipersistent time series, it was possible to establish a
relationship between spectral rigidity and antipersistence.  This
conjecture is supported by numerical calculations which involve atomic
nuclei and the classical Random Matrix Ensembles (RME)
\cite{Relano:02}. Later, theoretical expressions were derived in
Ref. \cite{Faleiro:04} using random matrix theory.  They explain,
without free parameters, the universal behavior of the excitation
energy fluctuations power spectrum.
The theory gives excellent agreement with numerical calculations. It
also reproduces the $1/f$ power law characteristic of chaotic systems and
the $1/f^2$ law characteristic of integrable system through several orders
of magnitude of the frequency interval.

All these results were obtained in the Random Matrix Theory (RMT) and
Nuclear Shell Model frameworks. These models allow an accurate
statistical analysis since very long sequences can be obtained with
high precision; moreover, RMT is the usual theoretical model in
studying quantum chaos, and the high excited states of the atomic
nuclei follow RMT predictions almost perfectly\cite{Haq:82,Gomez:04}.
However, other quantum chaotic systems cannot provide such quality of
data. Experiments usually provide short level sequences with some
precision problems: if two resonances are separated by a distance
smaller than the experimental line width, they are registered as one;
in microwave billiard experiments, certain resonances are missing
if the emitting antenna is positioned close to a nodal line.
Quantum systems with a
classical analogue can provide very long sequences, but they present
some non-universal features that spoil the universal behavior of
long-range correlations \cite{Berry:85,Graf:92}, which reveal some
dynamic properties of the classical analogue --the behavior of the
shorter periodic orbits of the system.  Therefore, analyzing whether
experimental limitations and the existence of short periodic orbits
modify our previous result becomes an important issue.

In this Letter we study the spectral fluctuations of experimentally
modeled quantum Sinai billiards. We find that the excitation energy
level fluctuations of these systems are in clear agreement with the
theoretical expressions found in Ref. \cite{Faleiro:04}. Therefore,
they exhibit $1/f$ noise, in spite of small deviations due to
experimental limitations and non-universal properties due to short
periodic orbits. Moreover, we present a simple analysis of these
anomalies using periodic orbit theory.

%
%

In the present work, quantum billiards have been substituted by
suitably shaped microwave resonators, whose eigenfrequencies are
measured. This kind of experiment uses the fact that the
time-independent Schr\"odinger equation and the wave equation are
mathematically equivalent. The data we analyze in this work has
already been used in Refs. \cite{stockprl,stockanp}, although in a
different context.
\begin{figure}[h]
\begin{center}
\includegraphics{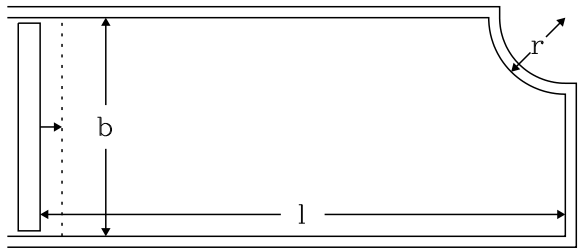}
\caption[]{Sketch of the quarter Sinai billiard with varying length
printed in scale.  The length of the billiard was varied from $l$ =
480\,mm to 460\,mm in steps of 0.2\,mm.  The width $b$ = 200\,mm and
the radius of the quarter circle $r$ = 70\,mm.}
\label{fig:sinaibil}
\end{center}
\end{figure}

To be specific, we have used a quarter Sinai billiard varying the
values of the length $l$ as depicted in Fig. \ref{fig:sinaibil}.  The
experimental setup makes it possible to measure resonances in the
frequency range 14.5 to 15.5 Ghz, where there are about 120
eigenvalues for each value of the parameter $l$. We have considered
about 100 of such sequences as realizations of a stochastic process
representative of the billiard, as we have previously done in
numerical experiments \cite{Relano:02}.

%
%

Prior to the statistical analysis, we have to remove the modulation
induced by the mean level density in the spectra. This procedure is
called unfolding; it consists in locally mapping the actual energy
levels $E_i$ into new dimensionless levels $\epsilon_i$
\begin{equation}
E_i \longrightarrow \epsilon_i=\overline{N}(E_i),\; i\in[1,\cdots,N],
\end{equation}
where $N$ is the dimension of the spectrum and $\overline{N}(E)$ is a
smooth approximation to the actual accumulated density. Quite
generally, $\overline{N}(E)$ can be calculated with the aid of the
following rule: in the limit of large E, the average density of
eigenstates is proportional to the volume of the phase space with
energy less than E \cite{Gutzwiller:90}. For quantum billiards we can
use an improved version of this result, which is also valid near of
the ground-state \cite{Graf:92}:
\begin{equation}
\overline{N}(E) = \overline{N}^{Weyl}(E)+\overline{N}^{bb}(E).
\label{Avg_dens}
\end{equation}
$\overline{N}^{Weyl}$ is the so called Weyl law; it includes a main
term related to the volume of the phase space, and two correction
terms related to the surface and the curvature of the boundary
\cite{Bloch:70}.  $\overline{N}^{bb}(E)$ is a further correction term
which takes into account the smooth trend produced by a set of
non-isolated marginally stable periodic orbits bouncing between the
two straight sides of the billiard.
%
%

After performing the unfolding procedure, we can calculate the nearest
neighbor level spacings defined by $s_i=\epsilon_{i+1} -
\epsilon_i,\,\,\,i\in [1,\cdots,N-1]$; since the unfolded spectrum has
a mean level density equal to 1, their average value is
$\left<s\right>=1$.  To study the long-range correlation structure, we
use the $\delta_n$ statistic \cite{Mehta:91,Relano:02} defined by
\begin{equation}
\delta_n  =  \sum_{i=1}^n (s_i - \left<s \right>)  = \varepsilon_{n+1}-\varepsilon_1-n,\;\;
n\in [0,\cdots,N-1].\\
\label{eq:delta}
\end{equation}
This statistic represents the deviation of the excitation energy of
the {\it (n+1)}th unfolded level from its mean value {\it n}, and it
is closely related to the spectrum fluctuations. Indeed, we can write
$\delta_n=-\widetilde{N}(E_{n+1})$ if we appropriately shift the
ground state energy; thus, $\delta_n$ represents the cumulated level
density fluctuations. If we consider the index $n$ as the analogue of
a discrete time \cite{aclaration}, we can analyze $\delta_n$ with
numerical techniques normally used in the study of complex
systems. The most simple is the analysis of the power spectrum
obtained from the discrete Fourier transform (DFT) \cite{DFT:81} of
the signal

\begin{equation}
\widehat\delta_k = \frac{1}{\sqrt{N}}\sum_{n=0}^{N-1} \delta_n
\exp\left(-\frac{2\pi i k n}{N}\right),\quad k \in[0,\cdots,N-1],
\end{equation}
as

\begin{equation}
S\left(\omega_k\right) = \left| \widehat{\delta}_k \right|^2,
\end{equation}
where
\begin{equation}
\omega_k = \frac{2\pi k}{N\Delta}
\label{freq_set}
\end{equation}
and  $\Delta$ is the sampling interval of the signal.

Here we face the following problem. The experiment cannot provide all
energy levels in the interval 14.5 and 15.5 GHz: there are a
few missing eigenvalues due to nodal lines of the wavefunction at the
position of the antenna.
Nevertheless, due to the level dynamics measurement, one can detect
where a certain level is missing. A comparison with eq. \eqref{Avg_dens}
shows that all levels are accounted for. Thus,  we are able to
arrange the known energy levels as $\{\varepsilon_{p_1},
\varepsilon_{p_2}, \cdots,\varepsilon_{p_i}, \cdots,
\varepsilon_{p_M}\}$, where $M (\le N)$ represents the total number of
levels whose energy is known, and $\{p_1,p_2,\cdots,p_M\}$ are their
positions in the spectrum. Therefore, the $\delta_n$ function can only
be calculated for certain whole numbers. Taking $\varepsilon_{p_1}$ as
a fictitious ground state, we have

\begin{equation}
\delta_{n_j}=\varepsilon_{p_{j+1}}-\varepsilon_{p_1}-n_j, \;\;
n_j = p_{j+1}-p_1,\;\;\; j\in[0,\cdots,M-1],\\
\end{equation}
Consequently, $\delta_{n_j}$ is not a standard time series, but an
unevenly spaced time series.

There are some obvious ways to get an evenly sampled signal from the
unevenly spaced $p_j$ points. Interpolation is one possible way, but
generally speaking, interpolation techniques perform poorly.
Therefore, we have decided to proceeded in a different way to analyze
the experimental data. For an arbitrarily sampled time series the
power spectrum can be estimated by the periodogram, $S(\omega) = \left|
\widehat{\delta}(\omega) \right|^2$, where $\widehat\delta(\omega)$ is
given by

\begin{equation}
\widehat\delta(\omega)=\frac{1}{\sqrt{M}}\sum_{j=0}^{M-1} \delta_{n_j}
\exp\left(-i\omega n_j\right).
\label{DFT}
\end{equation}

Eq. \eqref{DFT} is defined for any frequency $\omega$. However, since
the original signal is discrete, one should be able to chose a finite
set of frequencies to evaluate the Fourier transform, and then using
the values of eq. \eqref{DFT} at these frequencies, recover the
original signal. For evenly spaced time series of length $N$
eq. \eqref{freq_set} provides a {\it natural} set of frequency
values. This set has been built from a fundamental frequency $\omega_1
= 2 \pi /(N\Delta)$, which defines a sine wave of period equal to the
whole interval length; the remaining frequencies are multiples of
it. As commented above, the set $\{\widehat\delta(\omega_k)\}$, where
$k\in [0,1,\cdots,N-1]$, is known as the Discrete Fourier transform of
the signal.


In our case we have generated an appropriate set of frequencies
as follows:

(i) Since we know the smooth part of the level density (even if there
are some missing levels), we can perform the unfolding of the
spectrum, leading to a constant average level density. Thus, we can
set the fundamental sampling interval to $\Delta=1$.

(ii) Given a window ${\cal N} \le N$, we select all the experimental
billiards for which there is at least a pair of measured levels at
positions $p_a$ and $p_b$, satisfying that ${\cal N} = p_b - p_a$.

(iii) We consider the following frequency set $\omega_k = 2\pi k
/{\cal N}$ with $k= 1,2,\cdots$. This is probably the best choice
because the fundamental frequency, $\omega_1 = 2 \pi /{\cal N} $,
corresponds to a sine wave of period equal to the whole energy
interval $\varepsilon_{p_b} - \varepsilon_{p_a} \simeq {\cal N} \simeq
p_b -p_a.$ (Remember that after the unfolding procedure $<s>=1$.)

Due to the definition of our periodogram and to the selected frequency
set, we have that $S(\omega+2\pi) = S(\omega-2\pi) = S(\omega)$. For
this reason the power spectrum is defined in the basic interval
$[0,\pi)$. For evenly sampled series the frequency $\omega=\pi$ is
known as the Nyquist frequency and a change of trend in the power
spectrum can be observed near this upper limit.

(iv) Finally, the average power spectrum of the $\delta_n$ signal is
calculated using the selected billiards, and the experimental results are
compared with the theoretical expression found in
Ref. \cite{Faleiro:04}. Assuming that ${\cal N} \gg 1$, and that the
statistical properties of the spectral fluctuations are those of the
Gaussian orthogonal ensemble (GOE), the standard model for
time-reversal invariant systems, the universal behavior of the
$\delta_n$ power spectrum is given by

\begin{equation}
\avg{S(\omega_k)} =  \dfrac{K\lbr \fkn \rbr-1}{\omega_k^2}+
              \dfrac{K\lbr 1-\fkn \rbr-1}{(2\pi-\omega_k)^2}
              +\dfrac{1}{4\sin^2\lbr\dfrac{\omega_k}{2}\rbr}+
              \Theta,\;\; \omega_k\in(0,\pi),
\label{S_theor}
\end{equation}
Here,
\begin{equation}
K_{GOE}(\tau)  =
\left\{
\begin{array}{l}
2\tau - \tau \log(2\tau+1), \tau \le 1 \\
2 - \tau \log(\frac{2\tau+1}{2\tau-1}), \tau > 1
\end{array}
\right.
\label{K_theor}
\end{equation}
is the spectral form factor of GOE, whose analytical
expression is well known \cite{Mehta:91}. The constant shift
$\Theta$ reflects that $\delta_n$ is not a continuous two-point
measure (defined for every energy interval), but a discrete function
(defined only for each energy level). For GOE $\Theta=-1/12$,
provided that the whole sequence of $\delta_n$ values is included in
the calculation of the power spectrum. In this case, we cannot
evaluate $\delta_n$ for all $n$, but only for some energy levels. This
feature entails that the value of $\Theta$ should be affected by the
number $M$ of levels whose energy is known.
When $\omega_k \ll \pi$, eq. \eqref{S_theor} reduces to

\begin{equation}
\avg{S(\omega_k)} = \dfrac{1}{\pi \omega_k}, \;\; \omega_k \ll \pi,
\label{S_1f}
\end{equation}
showing that for small frequencies the excitation energy fluctuations
exhibit $1/f$ noise in GOE-like systems.


\begin{figure}[h]
\begin{center}
\includegraphics[height=9.5cm,angle=-90]{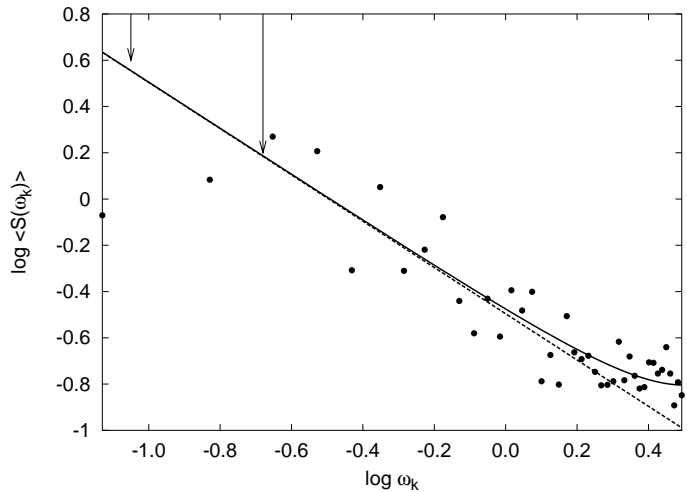}
\caption[]{{\bf Averaged power spectra of the $\delta_n$ function
for the Sinai microwave billiards} (filled circles) compared with the
theoretical prediction (solid line) as given by eq. \eqref{S_theor}.
The dashed straight line is the $1/f$ noise predicted by the
approximated expression \eqref{S_1f}. Vertical arrows indicate the
frequencies of the two shortest periodic orbits of the billiard.}
\label{fig:powerspectrum}
\end{center}
\end{figure}

In the forthcoming analysis we have chosen ${\cal N}=85$ since it is
large enough to obtain significant results, and it is possible to find
measured levels at positions $p_a$ and $p_b$ fulfilling $p_b - p_a =
85$ in almost every sequence of levels; in fact there are 97 sequences
satisfying this condition. Using a log-log scale,
Fig. \ref{fig:powerspectrum} compares the average power spectrum of
$\delta_n$ and the theoretical predictions of eq. \eqref{S_theor} with
$\Theta \approx -1/38$. This value was obtained by means of a
least-squares fit to the numerical data. In spite of the experimental
limitations, the agreement between the numerical and experimental
results is very good. Moreover, it is clearly seen that the spectral
fluctuations of these systems show a $1/f$ power law.
The significant power damping observed at lower
frequencies can be understood by a simple analysis of the contribution
of the bouncing ball orbits to the power spectrum. If we consider
$<l>$ = 470 mm and $<b>$ = 200 mm as the average length of the
straight sides of the billiards, the frequencies corresponding to
these orbits are $\omega=0.09$ and $\omega=0.21$. These values define
approximately the interval where the power is significantly
diminished; therefore, we can conclude that this phenomenon is due to
the removal of the contribution of bouncing ball orbits to the level
density.  All this is consistent with the idea that long period
classical orbits give rise to an universal behavior for integrable and
ergodic systems regardless of the system specificities
\cite{Ozorio:84}. Only when the main contribution comes from the
scarce short periodic orbits, deviations from the universal behavior
might be observed.

In summary, we have studied the spectral fluctuations of quantum Sinai
billiards modeled experimentally . We have considered the $\delta_n$
statistic, that characterizes the fluctuations of the excitation
energy, as a discrete time series. Using about one hundred different
lengths of the straight sides of the billiard, the average power
spectrum of $\delta_n$ was calculated. Recall that performing an
average is very important to reduce the fluctuations and clarify the
secular trend of the $\delta_n$ function. The main result of this
analysis is that the spectral fluctuations of these systems exhibit a
clear $1/f$ noise through almost the whole frequency domain. We have
also compared the theoretical predictions of eq. \eqref{S_theor} for
GOE-like systems with these results, finding an excellent agreement
except in the lower frequency region. Nevertheless, a simple study
reveals that these deviations from the universal behavior predicted by
RMT are due to the removal of the bouncing ball orbits during the
unfolding procedure; therefore there is no contribution to the power
spectrum in the low frequency domain. Last but not least, we can say
that $\delta_n$ statistic is not affected by the experimental
limitations, therefore we can conclude that it is a quite strong
statistic.

Prior to this work, we have found $1/f$ noise in energy spectrum
fluctuations for random matrix ensembles and atomic nuclei; these
results allowed us to conjecture that $1/f$ characterize spectral
fluctuations of chaotic quantum systems. The new results obtained in
this Letter corroborate this conjecture and give us a hint on the
non-universal features that can appear in the power spectrum at lower
frequencies due to the short periodic orbits.


This work is supported in part by Spanish Government grants
BFM2003-04147-C02 and FTN2003-08337-C04-04.
The experiments have been supported
by the Deutsche Forschungsgemeinschaft via several grants.

\end{document}